# برهمکنش های فوق ریز در بلور $USb_2$

فتحی، آرش [۱و۲]؛ جلالی اسد آبادی ، سعید [۲و۳]؛ گشتاسبی راد،محمد [۱]؛امینی ، حامد [۲و۴]

[۱] گروه فیزیک دانشگاه سیستان و بلوچستان ،زاهدان

[۲] گروه فیزیک دانشگاه اصفهان ، اصفهان

[۳] مرکز پژوهش دانش و فناوری نانو، دانشگاه اصفهان ، اصفهان

[۴] انجمن علمی دانشجویی گروه فیزیک دانشگاه اصفهان ، اصفهان

## چکیده

در این پژوهش برهمکنش های فوق ریز در مکان اورانیوم از ترکیب آنتی فرومغناطیس $USb_2$ با استفاده از نظریه تابعی چگالی و روش امواج تخت بهبود یافته بعلاوه اوربیتالهای موضعی ($APW + lo$) بررسی شده است. وابستگی برهمکنش چهارقطبی هسته ای به ساختار مغناطیسی در این ترکیب مورد بررسی قرار می گیرد. نتایج نشان می دهد که الکترونهای$5f$ تمایل دارند که به خوبی با الکترون های رسانش هیبرید شوند.

# Hyperfine Interaction in USb$_2$ Crystal

**Fathi, Arash[1,2] ; Jalali Asadabadi, Saeid[2,3] ;Goshtasbi Rad,Mohammad[1]; Amini, Hamed[2,4]**

[1]*Physics Department, University of Sistan & Baluchestan, Zahedan*
[2]*Physics Department, University of Isfahan Isfahan*
[3]*Research center for nano sciences and nano technology , The University of Isfahan, Isfahan*
[4]*Student Scientific society of physics Dept. of Isfahan University, Isfahan*

## Abstract

*The hyperfine interactions at the uranium site in the antiferromagnetic USb$_2$ compound were calculated within the density functional theory (DFT) employing augmented plane wave plus local orbital (APW+lo) method. We investigated the dependence of the nuclear quadruple interaction to the magnetic structure in USb$_2$ compound. The result shows that the 5f-electrons have the tendency to be hybridized with the conduction electrons.*

PACS No. 75

## مقدمه

نظم اوربیتالی برای اولین بار به روش پراکندگی تشدیدی $X-Ray$ مشاهده شد و از آن پس آزادی اوربیتالی به عنوان یک پارامتر مهم در فیزیک حالت جامد شناخته شده است.[1] چنین نظمی در بعضی از ترکیبات عناصر خاک نادر و آکتنیدها که به آن نظم چهارقطبی در دستگاه های الکترونی$4f$ و$5f$ گوییم، با روش های مشابهی قابل مشاهده است.

برهمکنش چهار قطبی هسته ای که بر همکنش میان یک بخش الکترونی به نام عملگر تانسور گرادیان میدان الکتریکی و یک بخش هسته ای به نام عملگر تانسور چهارقطبی الکتریکی است می تواند در مشخص کردن توزیع گشتاور چهارقطبی الکتریکی در دستگاه های $f$ الکترونی و دستگاه های جایگزیده به کار رود.[2,3]

از آنجا که سهم گشتاور چهارقطبی هسته ای در هسته ی $^{238}U$ مقداری قابل توجه است (3.2– barns)، برهمکنش چهارقطبی هسته ای می تواند ابزاری مفید در توضیح چگونگی توزیع چهارقطبی های $5f$ باشد.

$USb_2$ که ماده ای آنتی فرو مغناطیس با دمای نیل بزرگ نزدیک به دمای اتاق است، یکی از ترکیبات سری $UX_2 (X = As, Sb, P\ and\ Bi)$ محسوب می شود. این ترکیب

دارای ساختار تتراگونال از نوع $Cu_2-Sb$ با گروه فضایی ($P_4\,nmm$) است. گشتاورهای مغناطیسی اتم های U در این ترکیب همگی در جهت [001] در راستای عمود بر صفحات (001) و در یک دنباله به صورت (↑↓↓↑) قرار دارند. یاخته بسیط در این ترکیب دو برابر یاخته بسیط شیمیایی است که در شکل(۱) آورده شده است.

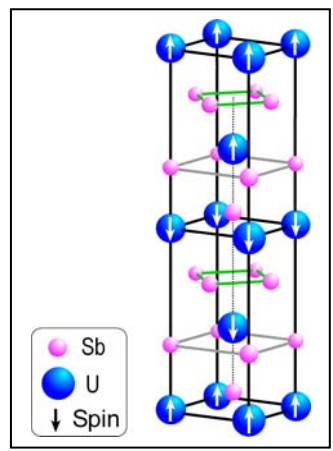

شکل(۱): چینش اسپینی (↑↓↓↑) در یاخته مغناطیسی $USb_2$

## روش تحقیق و نحوه ی محاسبات

در این پژوهش به محاسبه گرادیان میدان الکتریکی در محل اتم U و همچنین محاسبه برهمکنشهای فوق ریز در بلور $USb_2$ پرداخته ایم. محاسبات ما بر اساس نظریه تابعی چگالی (DFT) و به روش پتانسیل کامل امواج تخت بهبود یافته بعلاوه اربیتالهای موضعی الکترونهای ظرفیت ($FP-APW+lo$) و با اعمال تقریب شیب تعمیم یافته (GGA) بر پانسیل تبادلی همبستگی، با استفاده از کد $WIEN2k$ می باشد[4]. زاویه بین محور اصلی گرادیان میدان الکتریکی (EFG) و میدان مغناطیسیِ فوق ریز ($H_{hf}$) می تواند مقادیر بین صفر تا نود درجه را اختیار کند. ساختار مغناطیسیِ $USb_2$ با در نظر گرفتن جهت گشتاور مغناطیسی، به وسیله آزمایشهای پراکندگی نوترون تعیین می شود[5]. از آنجا که گشتاور مغناطیسی در $USb_2$ خود را در امتداد محور c تطبیق می دهد، جهت ($H_{hf}$) نیز در امتداد محور c است. به عبارت دیگر از آنجایی که ساختار بلوری $USb_2$ به صورت یک تتراگونال با یک یاخته بسیط کشیده شده در امتداد

محور c و شامل لایه های متناوبی از اتم های U و X است، می توان فرض کرد که محور اصلیِ تانسور گرادیانِ میدان الکتریکی، یا در امتداد محور c است و یا در داخل صفحه ab در حالت آنتی فرو مغناطیس قرار دارد.

هنگامی که تانسور (EFG) موازی محور c قرار می گیرد زاویه بین محور اصلی تانسور (EFG) و ($H_{hf}$) صفر است و هنگامی که در صفحه ab قرار دارد این زاویه برابر نود درجه می باشد. نتایج نشان می دهد که در این ترکیب، برهم کنش چهارقطبی هسته ای برای حالت $\theta = 0°$ منفی و برابر ($-35 \pm 9\ \ mm s^{-1}$) است؛ این مقدار در حالت $\theta = 90°$ همان مقدار حالت $\theta = 0°$ است با این تفاوت که در یک عدد ($-2$) ضرب شده است.

پارامترهایِ فوق ریز برای $\theta = 0°$ در جدول (۱) داده شده اند.

| تـرکیب | $e^2qQ$ | EFG ($Vnm^{-2}$) | $H_{hf}$ (T) | $A_{hf}$ ($T/\mu\beta$) |
|---|---|---|---|---|
| $USb_2$ | $-35 \pm 9$ | $1.6 \pm 0.4 \times 10^4$ | $270 \pm 20$ | $50 \pm 10$ |

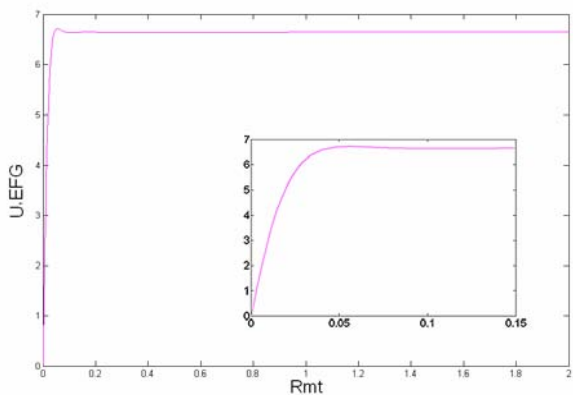

جدول(۱): بر هم کنش های فوق ریز در $USb_2$
شکل(۲): نمودار گرادیان میدان الکتریکی بر حسب شعاع کره موفن تین

همچنین نمودار گرادیان میدان الکتریکی بر حسب شعاع کره موفن تین برای اتم U در شکل(۲) آورده شده است. این

نتایج کار حاضر نشان می دهد که بر همکنش چهار قطبی هسته ای به ساختار مغناطیسی ترکیبات $UX_2$ بستگی دارد. به طور کلی تانسور گرادیان میدان الکتریکی (EFG) به چهارقطبی های 5f، همچنین به نظم اتمی جایگزیده (سهم شبکه ای) و به الکترونهای رسانش وابسته است. ما فرض می کنیم که سهم الکترونهای رسانش در تانسورِ (EFG) کوچک باشد. حال انتخاب این که از بین دو سهم تجربی باقی مانده کدام یک بر دیگری برتری دارد، کار مشکلی است به همین خاطر معمولاً در کارهای تجربی نقش هر دو سهم یکسان در نظر گرفته می شود، اما بر طبق محاسبات ما سهم غالب در برهم کنشهای چهارقطبی هسته ای ناشی از چهارقطبی های 5f است.

از ثابت جفت شدگی فوق ریز بدست آمده می توان این طور نتیجه گرفت که چهارقطبی 5f و الکترونهای رسانش در این ترکیب به خوبی هیبرید شده اند، چرا که مقدار ثابت جفت شدگی فوق ریز ($A_{hf}$) به طور کلی وابسته به هیبریدشدگی الکترونی بین اتمهای مغناطیسی و اتمهای لیگاند است. ثابت جفت شدگی فوق ریز، وابسته به بزرگیِ گشتاور مغناطیسی و توزیعی از توابع موج بیرون اتمهای اورانیوم است. ثابت جفت شدگی در $USb_2$ برابر با مقدار آن در بیشتر ترکیبات غیر فلزی U است.

## نتیجه گیری

در این پژوهش برهمکنش های فوق ریز در حالت آنتی فرو مغناطیس در ترکیب $USb_2$ بررسی شده است. برهمکنش چهار قطبی هسته ای به رشته های مغناطیسی در این دستگاه وابسته است. به نظر می رسد گوناگونی برهمکنش های چهارقطبی هسته ای در میان ترکیبات $UX_2$ نتیجه ای از سهم چهار قطبی های 5f باشد. این نشان می دهد که واقعاً الکترونهای 5f تا حد زیادی با الکترونهای رسانش هیبرید می شوند.

## سپاسگزاری

## مراجع